\documentclass[aps,prb,twocolumn,groupedaddress]{revtex4-1}

\newcounter{seqncounter}
\newcounter{sfigcounter}
\usepackage{mathrsfs,amsmath} 
\usepackage{amsfonts}
\usepackage{amssymb}
\usepackage{fixmath}
\usepackage{graphicx}
\usepackage{epstopdf}
\usepackage{setspace}
\usepackage[toc,page]{appendix} 
\usepackage{hyperref,hypcap}
\usepackage{placeins}
\usepackage{chngcntr}
\usepackage{titlesec}
\usepackage{cancel}
\usepackage{enumerate}
\usepackage{slashed}
\usepackage{tikz}

\begin{document}


\title{Shear sound of two-dimensional Fermi liquids}

\author{Jun Yong Khoo}
\affiliation{Department of Physics, Massachusetts Institute of Technology, 77 Massachusetts Avenue, Cambridge, Massachusetts 02139, USA}
\author{Inti Sodemann Villadiego}
\affiliation{Max-Planck Institute for the Physics of Complex Systems, D-01187 Dresden, Germany}


\date{\today}

\begin{abstract}
We study the appearance of a sharp collective mode which features transverse current fluctuations within the bosonization approach to interacting two-dimensional Fermi liquids. This mode is analogous to the shear sound modes in elastic media, and, unlike the conventional zero sound mode, it is damped in weakly interacting Fermi liquids and only separates away from the particle-hole continuum when the quasiparticle mass becomes twice the transport mass $m^* \gtrsim 2 m$. The shear sound should be present in a large class of interacting charged and neutral Fermi liquids especially those proximate to critical points where the quasiparticle mass diverges. In metals this mode remains linearly dispersing in the presence of the long-ranged Coulomb force, unlike the conventional zero sound mode which becomes the plasma mode. We also detail a quick path between bosonization and classical Landau's Fermi liquid theory by constructing a mapping between the solutions of the classical kinetic equation and the quantized bosonic eigenmodes. By further mapping the kinetic equation into a 1D tight-binding model we solve for the entire spectrum of collective and incoherent particle-hole excitations of Fermi liquids with non-zero $F_0$ and $F_1$ Landau parameters.
\end{abstract}


\maketitle


\section{Introduction}
When Landau introduced his theory of the Fermi liquid more than 60 years ago~\cite{Landau1,Landau2} it was not immediately clear the extent to which it was an approximate description. Subsequent developments, such as the theorem of Luttinger~\cite{Luttinger} asserting the adiabatic invariance of the Fermi volume, contributed to strengthen the belief on the essential validity of Landau's theory. At the dawn of the twentieth century the advent of modern approaches like the renormalization group of fermions~\cite{Shankar} and higher dimensional bosonization~\cite{Luther,Haldane,Marston,Fradkin} contributed to cement the agreement that in two-dimensions and higher Landau's Fermi liquid theory (LFLT) captures the essential long wavelength and low energy behavior of a large class of interacting systems with a Fermi surface known as Landau Fermi liquids (LFL).

\begin{figure}
\centering
\includegraphics[scale=0.33]{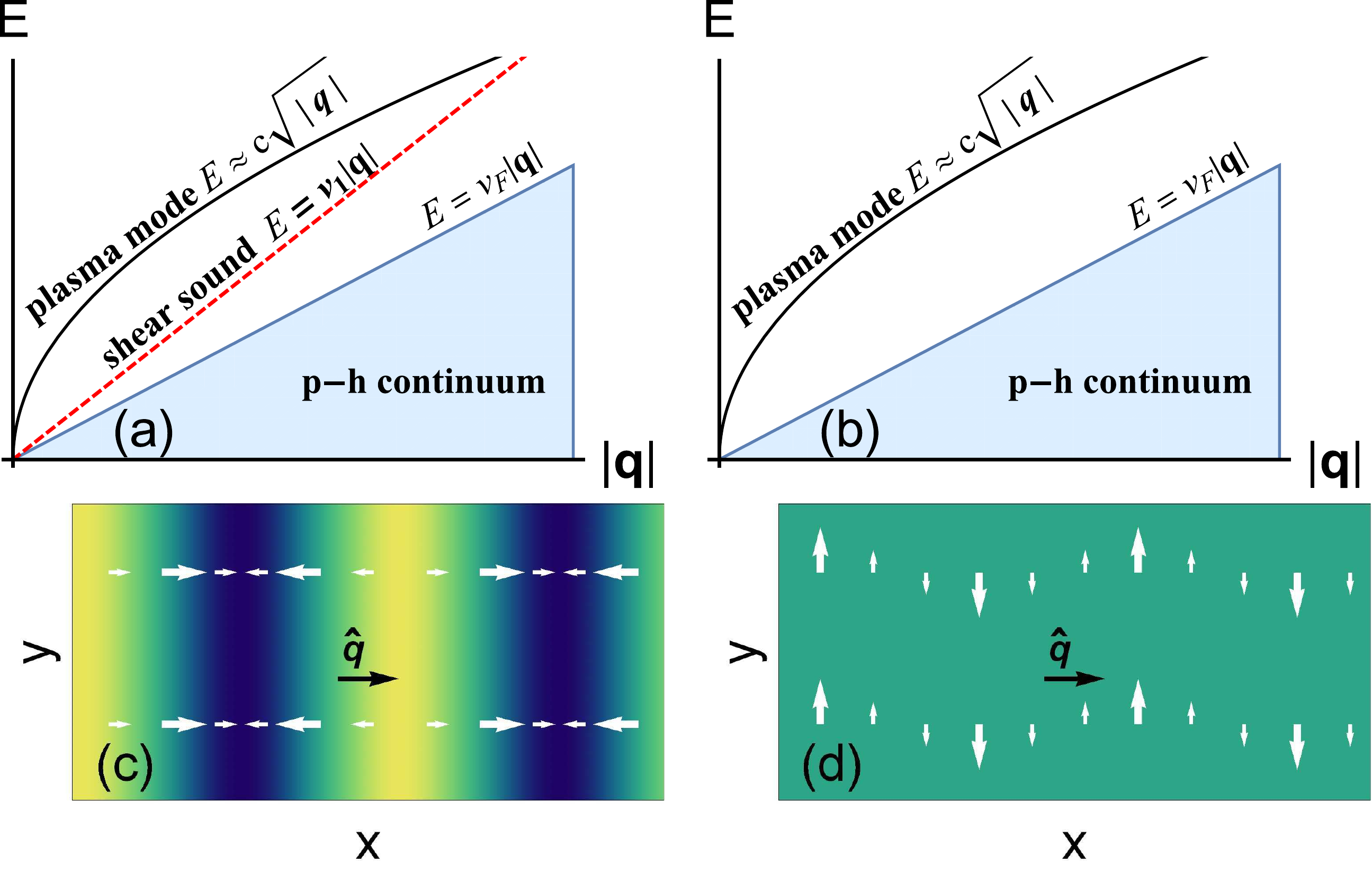}
\caption{\label{cartoon}Schematic of the particle-hole (p-h) continuum and the collective modes of the Fermi liquid in 2D (a) with and (b) without shear sound (red, dashed line). In the case of neutral Fermi liquids, the zero sound (plasma modes in metal) will have a linear dependence instead of the square root dependence shown here. (c) Zero and (d) shear sound with wavevector ${\bf q}$ parallel to the $x$-axis. The color scale represents the density and the arrows the current fluctuations.}
\end{figure}

Typically weakly interacting LFL have a single sharp collective excitation known as the zero sound mode, with the remaining excitations forming incoherent particle-hole continuum~\cite{Pines,Baym}. This mode becomes the plasma mode in charged LFLs as depicted in Fig.~\ref{cartoon}(a) due to the presence of the long range Coulomb interactions. However, as we will demonstrate, a well separated shear sound mode can emerge from the particle-hole continuum and become an additional sharp collective mode in LFLs once the interactions are beyond certain threshold. This unconventional collective mode has the characteristics of a {\it shear sound} wave resembling the transverse excitations of an elastic medium, which can oscillate transversely to the propagation direction of the shear wave as depicted in Fig.~\ref{cartoon}(d). This mode is absent in classical fluids due to their vanishing shear modulus and its presence in the quantum Fermi liquids is a vivid reminder of the deep differences between quantum and classical fluids. 

The possibility of a shear sound mode in interacting LFLs was long ago recognized by the $^3$He community~\cite{Khalat,Fomin,Pines,Baym,Lea}. However, in 3D Fermi liquids the mode requires typically a larger interaction strength to separate from the particle-hole continuum and it also remains closer to the particle-hole continuum than in 2D. Although there is evidence for the presence of this mode in three-dimensional $^3$He~\cite{Roach}, its quantitive understanding remains elusive~\cite{Flowers}, largely because of its proximity to the particle-hole continuum. As we will argue, this mode should be easier to observe in two-dimensions and might be present in a variety of experimentally accessible metallic and neutral Fermi liquids. Approximately, we expect the mode to separate from the particle-hole continuum once the quasiparticle mass is renormalised beyond a factor of two compared to the bare mass.

To study this unconventional collective mode, we apply the higher dimensional bosonization formalism to LFLs. This formalism is essentially a second quantized field theoretic version of Landau Fermi liquid theory. We will describe a short path between the bosonization and conventional Fermi liquid theory which highlights this intimate connection. In particular, the harmonic nature of the bosonized theory leads to an equivalence between classical and quantum equations of motion for collective modes, analogous to how the Ehrenfest theorem relates the dynamics of classical and quantum Harmonic oscillators.  Therefore, for purposes of collective modes, the results are equivalent to those of classic LFLT~\cite{Luther,Haldane,Marston,Fradkin}. Our approach is inspired by and in close connection to recent {\it top-down} approaches to bosonization which start from LFLT viewed as a classical field theory and construct from it a quantum field theory by inferring the quantization relations of its classical variables~\cite{Senthil,Son}.

\section{Formalism}
The long-wavelength and low-energy description of most phases of matter involves a finite number of continuum fields related to conservation laws and order parameters in the case of broken symmetry phases. However, LFL depart radically from this, in that they have an {\it infinite} number of slow degrees of freedom which parametrize the shape of the Fermi surface~\cite{Haldane}. The state of a LFL can be parametrized by the Fermi radius at any point in space ${\bf x}$, $p^{\text{F}}_{{\bf x},\theta} = p_{\text{F}}^0 + u_{{\bf x},\theta}$, where $\theta$ is the angle on the Fermi surface. In bosonization the Fermi radius becomes a quantum mechanical operator whose algebra is given by~\cite{Haldane,Marston,Fradkin,Senthil,Son}:

\begin{eqnarray}\label{ucomm}
\left[\hat{u}_{{\bf x},\theta} , \hat{u}_{{\bf x}',\theta'}\right] =\frac{(2\pi)^2}{i p_{\text{F}}} \delta (\theta  - \theta ') \partial _{n}\delta({\bf x}-{\bf x}') + O(\hat{u}),
\end{eqnarray}

\noindent where $ \partial _{n}=\hat{{\bf p}}_\theta \cdot \partial_{{\bf x}} $ is the derivative along the normal $\hat{{\bf p}}_\theta$ of the Fermi surface. We introduce a matrix notation for $\theta$ that will compactify our formulas, by defining:

\begin{eqnarray}
v_\theta^{\dagger} G_{\theta,\theta'} w_{\theta'} \equiv \int  d\theta d\theta ' v^*(\theta) G(\theta,\theta') w(\theta') .
\end{eqnarray}

\noindent With this notation, the Hamiltonian governing the dynamics of the Fermi surface can be written as:

\begin{eqnarray}\label{Ham}
\hat{H} =\int d^2{\bf x}  \ \hat{u}_{{\bf x},\theta}^\dagger h_{\theta,\theta'}  \hat{u}_{{\bf x},\theta'},
\end{eqnarray}

\noindent where $h(\theta,\theta')=v_{\text{F}} p_{\text{F}}(2 \pi \delta(\theta' - \theta )+F(\theta' - \theta ))/2(2\pi)^3$. $F(\theta' - \theta )$ is the Landau function characterizing the interactions between quasiparticles~\footnote{We focus on circularly symmetric Fermi surfaces and note that the Fermi radius field $\hat{u}_{{\bf x},\theta}$ is understood to be a Hermitian operator.}.
Notice that LFLT has an infinite number of conserved quantities which measure the spatially averaged shape of the Fermi surface. Formally, any operator of the form $\hat{g}(\theta)= \int d^2{\bf x} \ g(\theta) \hat{u}_{{\bf x},\theta}$ is a conserved quantity.

To exploit translational invariance we introduce the Fourier modes of the Fermi surface deformations $\hat{u}_{{\bf q},\theta}\equiv \int d^2{\bf x}  \ \hat{u}_{{\bf x},\theta} e^{-i{\bf q}\cdot {\bf x}}$. These operators can be interpreted as {\it bare} particle-hole creation operators  $c^\dagger_{{\bf p}+{\bf q}/2} c_{{\bf p}-{\bf q}/2}$ with ${\bf p}$ coarse grained over a region near the angle $\theta$ on the Fermi surface~\cite{Fradkin,Fradkin2}. The equation of motion following from Eqs.~\eqref{ucomm} and \eqref{Ham} for these operators is:

\begin{eqnarray}
i \partial_t \hat{u}_{{\bf q},\theta} &=&\left[\hat{u}_{{\bf q},\theta}, \hat{H} \right]=K_{\theta, \theta'} \hat{u}_{{\bf q},\theta'}, \\
K(\theta, \theta') &=& v_{\text{F}} {\bf q} \cdot \hat{{\bf p}}_\theta \left(\delta(\theta - \theta' )+\frac{1}{2 \pi}F(\theta - \theta' )\right).
\end{eqnarray}

\noindent The equation above can be recognized to be an operator version of the classic Landau's linearized kinetic equation~\cite{Pines,Baym}.
Notice that $\hat{u}_{{\bf q},\theta}$ do not satisfy canonical bosonic commutation relations and that the kinetic matrix, $K_{\theta, \theta'}$, is non-Hermitian. However, there exists a simple similarity transformation between $K$ and its Hermitian conjugate:

\begin{eqnarray}
K = T K^\dagger T^{-1}, \ \ T_{\theta, \theta '} =  \frac{(2\pi)^2{\bf q} \cdot \hat{{\bf p}}_\theta}{p_{\rm F}} \delta(\theta - \theta').
\end{eqnarray}

\noindent We are now in a position to state a mapping between the classical solutions of Landau's kinetic equation and their quantum counterpart. For each {\it classical} eigenfunction of the kinetic equation, $\psi_{\lambda,{\bf q},\theta}$, there is a {\it quantum} eigenmode, $\hat{\psi}_{\lambda,{\bf q}}$, given by:

\begin{eqnarray}
\hat{\psi}_{\lambda,{\bf q}} = \psi_{\lambda,{\bf q},\theta}^\dagger T_{\theta, \theta '}^{-1} \hat{u}_{{\bf q},\theta'}.
\end{eqnarray}

\noindent where $K_{\theta, \theta'} \psi_{\lambda,{\bf q},\theta'}=E_\lambda \psi_{\lambda,{\bf q},\theta}$ and $i \partial_t \hat{\psi}_{\lambda,{\bf q}} =E_\lambda \hat{\psi}_{\lambda,{\bf q}}$. By choosing a suitable normalization for the classical solutions, $ \psi_{\lambda,{\bf q},\theta}^\dagger T_{\theta, \theta '}^{-1} \psi_{\lambda',{\bf q},\theta'} = {\rm sgn}(E_\lambda)\delta_{\lambda,\lambda'}$, we arrive at canonical bosonic eigenmodes describing the fluctuations of the shape of the Fermi surface:

\begin{eqnarray}\label{eq.psicomm}
\left[ \hat{\psi}_{\lambda,{\bf q}} , \hat{\psi}_{\lambda',{\bf q'}}^\dagger \right] = (2\pi)^2 \delta({\bf q}-{\bf q}') \text{sgn}(E_\lambda) \delta _{\lambda,\lambda'},
\end{eqnarray}

\noindent where the sign of the eigenvalue $E_\lambda$ dictates which one of the pair $\hat{\psi}_{\lambda,{\bf q}} , \hat{\psi}_{\lambda',{\bf q'}}^\dagger$ is the raising and which one is lowering operators. These eigenmodes describe both collective oscillations such as the zero sound and also the continuum of particle-hole excitations. Any two-body operator can be represented as a linear combination of these modes and in particular: $\hat{u}_{{\bf q},\theta}=\sum_{\lambda}\text{sgn}(E_\lambda) \hat{\psi}_{\lambda,{\bf q}} \psi_{\lambda,{\bf q},\theta}$.

\section{Mapping to a chain}
As we have seen, the quantum problem reduces to the eigenvalue problem of the classic kinetic equation. We begin by simplifying the classical eigenvalue problem by exploiting its symmetries. Rotational symmetry allows us to restrict ${\bf q}=q \hat{{\bf x}}$. We measure the angle along the Fermi surface, $\theta$, from this axis. Additionally, we assume a mirror symmetry $F(\theta) = F(-\theta)$, $K_{\theta, \theta'}=K_{-\theta, -\theta'}$, which decouples the even and odd parity eigenmodes, which we label with a superscript $\sigma=\pm$ denoting: $\psi^{\sigma}_{\lambda,{\bf q},\theta}=\sigma \psi^{\sigma}_{\lambda,{\bf q}, -\theta}$.

There is also a time-reversal symmetry $K_{\theta, \theta'}^*=K_{\theta, \theta'}$ which implies that the eigenfunctions can be taken to be purely real~\footnote{We are assuming the eigenvalues to be real, which requires the Hamiltonian in Eq.~\eqref{Ham} to be positive definite, which is the usual condition to avoid Pomeranchuk-like instabilities.}. The kinetic equation also has a particle-hole-like symmetry which follows from an inversion in momentum space: $K_{\theta+\pi, \theta'+\pi}=-K_{\theta, \theta'}$. Therefore the eigenfunctions, $\psi^{\sigma}_{\lambda,{\bf q},\theta}$, come in pairs with opposite eigenvalues. Namely, if $\psi^{\sigma}_{\lambda,{\bf q},\theta}$ is an eigenfunction with eigenvalue $E^{\sigma}_\lambda$, then $\psi^{\sigma}_{\lambda,{\bf q},\theta+\pi}$ is an eigenfunction with eigenvalue $-E^{\sigma}_\lambda$. For fixed ${\bf q}$ these two solutions describe physically distinct modes. The one with positive (negative) eigenvalue will be a creation (destruction) operator, and, its destruction (creation) operator partner will live in the space of excitations with momentum $-{\bf q}$. This feature can be traced back to the property that particle-hole excitations with small momentum $ {\bf q}$ can only be created in one of the halves of the Fermi surface satisfying ${\bf q} \cdot \hat{{\bf p}}_\theta>0$.

We describe now a convenient representation of the kinetic equation in a similar spirit to a recent treatment of spin orbit coupled systems~\cite{1dchain}. We begin by decomposing into angular momentum channels ($ {\bf q}$ implicit below):

\begin{eqnarray}
F(\theta) &=& F_0+ \sum_{l=1}^\infty 2 F_{l} \cos (l\theta), \\
\psi^{+}_{\lambda,\theta}
&=& \psi^{+}_{\lambda,0}+\sum_{l=1}^\infty 2 \psi^+_{\lambda,l} \cos (l\theta), \label{eq.psieven} \\
\psi^{-}_{\lambda,\theta}
&=& \sum_{l=1}^\infty 2 \psi^-_{\lambda,l} \sin (l\theta). \label{eq.psiodd}
\end{eqnarray}

\noindent With this the kinetic equation takes the form of a non-Hermitian tight-binding model in which the sites are the angular momentum channels:

\begin{eqnarray}\label{TBmodel}
E^{\sigma}_\lambda \psi^{\sigma}_{\lambda,l+1} = t _{l} \psi^{\sigma}_{\lambda,l} + t_{l+2} \psi^{\sigma}_{\lambda,l+2},  \label{eq.recrelpsi}
\end{eqnarray}

\noindent where $t _{l}=v_{\rm F}q (1+F_l)/2$.
The above equation applies when $l\geq 0$ for $\sigma=+$ and when $l\geq 1$ for $\sigma=-$ and is accompanied by the corresponding boundary conditions,

\begin{eqnarray}
E^{+}_\lambda \psi^{+}_{\lambda,0} &=& 2t _{1} \psi^{+}_{\lambda,1}, \label{initeven} \\
E^{-}_\lambda \psi^{-}_{\lambda,1} &=& t _{2} \psi^{-}_{\lambda,2}.\label{initodd}
\end{eqnarray}

We see that the Landau parameters play the role of bond-disorder in the effective tight binding model. Notice that the eigenvalue problem for the odd modes is completely independent of $F_0$. A remarkable property which becomes transparent in this way of writing the problem is that there exists a simple relation between the eigenvalue problem in the odd and even subspaces. Namely, the eigenvalue problem in the even sector for a set of Landau parameters $\{ F_l\}$ can be mapped into the problem in the odd sector with modified Landau parameters $\{ F'_l\}$ by relabelling sites as $l \rightarrow l+1$, such that the Landau parameters are related by $F'_{l+1}=F_{l}$, and by removing the factor of 2 when going from the even Eq.~\eqref{initeven} to odd boundary condition Eq.~\eqref{initodd}.

\section{Shear sound}
We begin by considering the simplest interacting Fermi liquid with only a non-zero s-wave Landau parameter, $F_0 \neq 0 $ and $F_{l > 0}= 0$. In this case the tight binding chain has only one defective bond connecting the $l=0$ site at the end of the chain in the even sector $\sigma=+$. As detailed in the Supplemental Material
~\footnote{See Supplemental Material for derivation of kinetic equation solutions and response functions.},
 one can solve Eq.~\eqref{TBmodel} recursively. There are two kinds of solutions. The first kind form the analog of a ``band" and describe excitations in the particle-hole continuum ($E<v_F q$), and are found to be (up to global constant):

\begin{eqnarray}\label{eq.psisolnosc}
\psi _{l\geq 1}^+ 
&=&\frac{\sin(l+1)\theta _{E} -\frac{E}{v_{\rm F}q} \sin l\theta _{E} -F_0\sin (l-1)\theta _{E}}{\sin \theta _{E}}, \nonumber\\
\end{eqnarray}

\noindent where $\cos \theta_E=E/v_{\rm F}q$ and parametrizes the angle on the Fermi surface where the particle-hole pair is created. The second kind are isolated solutions analogous to bound states created by the ``bond-disorder". The $F_0$ model has a single isolated bound state that is present only for $F_0>0$ and corresponds to the celebrated zero sound mode. Its dispersion is found to be~\cite{FL0}:

\begin{equation}\label{eq.v0}
\frac{E_0}{v_{\text{F}} q}=\frac{1+F_0}{\sqrt{1+2F_0}}, \  \ F_0>0, 
\end{equation}
And the wavefunction of the zero-sound is:
\begin{eqnarray}\label{eq.psi0wf}
\psi^+ _{l\geq 1} &=& \psi^+ _{0} \frac{1+F_0}{(1+2F_0)^{l/2}},\\
\psi^+ _{0} &=& \left(\frac{2\pi q}{p_{\rm F}} \frac{v_{\text{F}} q}{E_0} \frac{F_0 (1 + F_0)}{(1+2F_0)^2}\right)^{1/2}.
\end{eqnarray}

\noindent In the $F_0$ model the odd parity modes are identical to the non-interacting Fermi gas, and, hence there is no transverse collective modes and only the particle-hole continuum. We will now consider a more realistic model of the LFL which has non-vanishing $\{F_0,F_1\}$ Landau parameters. The mapping described in the previous section between odd and even parity sectors immediately implies that this model can support an undamped collective odd mode. The dispersion and wavefunction of this mode is found to be:

\begin{eqnarray}
\frac{E_1}{v_{\text{F}} q}& = &\frac{1+F_1}{2\sqrt{F_1}}, \quad F_1 > 1, \label{eq.v1} \\
\psi^- _{l\geq 2} &=&  \psi^- _{1} \frac{1+F_1}{F_1^{(l-1)/2}}, \ \psi^- _{1} = \left(\frac{\pi q}{8 p_{\rm F}} \frac{v_{\text{F}} q}{E_1} \left( 1 - \frac{1}{F_1^2} \right)\right)^{1/2} \label{eq.psi1wf}. \nonumber \\
\end{eqnarray}

\begin{figure}[h]
\centering
\includegraphics[scale=0.33]{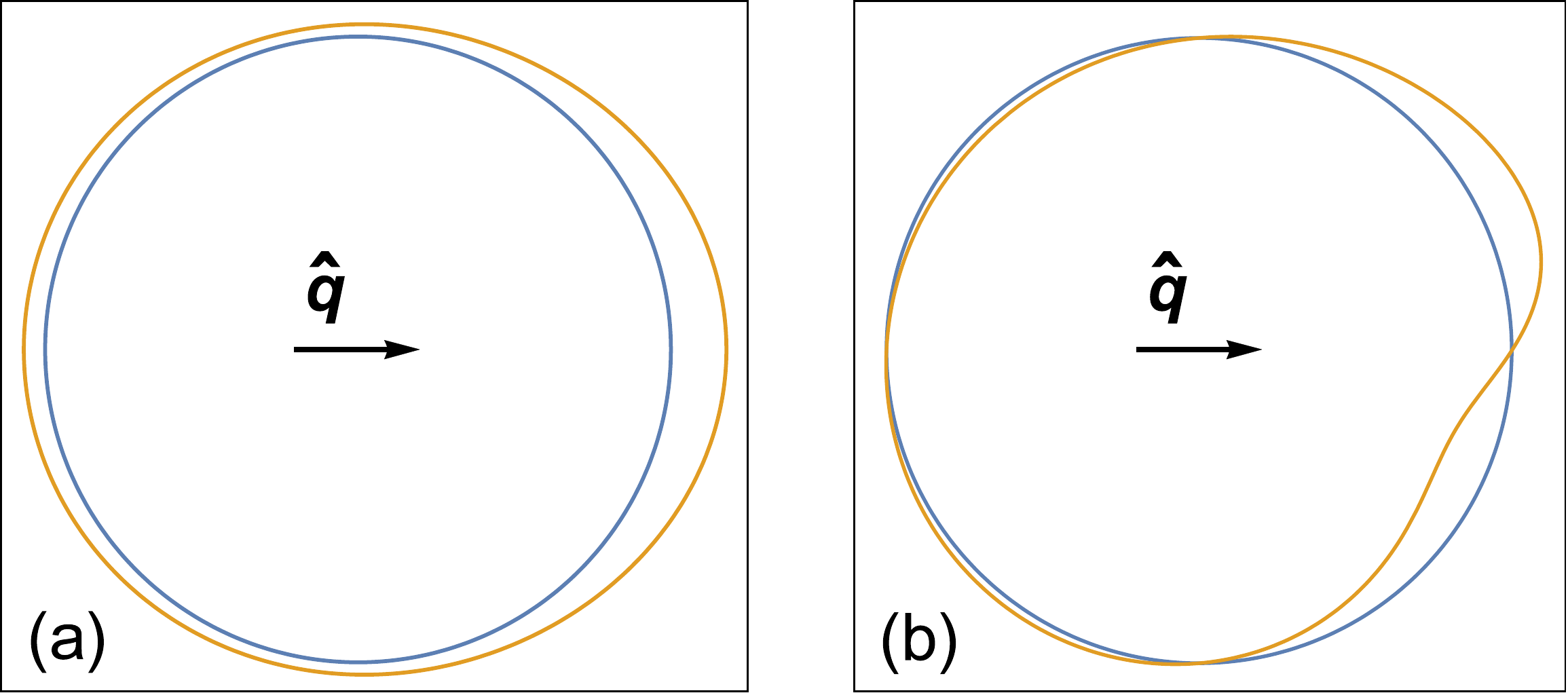}
\caption{\label{modepic} Fermi surface deformations for zero (a) and shear (b) sound eigenmodes ($F_0 = 1$, $F_1 = 3$).}
\end{figure}

\noindent The shape of the Fermi surface deformations associated with shear and zero sound modes are illustrated in Fig.~\ref{modepic}. As we will see this extra collective mode features transverse current fluctuations with no density oscillations in analogy with the shear sound of elastic media. The even sector gets modified by the introduction of a finite $F_1$, the details of which will be discussed in Sec.~\ref{sec:marker}. This modification is unessential to our current discussion for $F_0, F_1 > 0$.

The study of shear fluctuations of interacting electrons has an important precedent in the work of Conti and Vignale~\cite{Conti} (see also Ref.~\onlinecite{Vignale}).
Our expression for the shear sound velocity in Eq.~\eqref{eq.v1} is in agreement with theirs (see Eq.~(4.12) from Ref.~\onlinecite{Conti}) in the regime in which the collective mode is well separated from the particle hole continuum, namely when $\omega \gg v_{F} q$, which requires $F_1 \gg 1$~\footnote{Notice that in Ref.~\onlinecite{Conti} the results are expressed in terms of the bare Fermi velocity and we use a different convention for the $F_1$ Landau parameter, which is $F_{1,{\rm theirs}}= 2 F_{1,{\rm ours}}$. We are thankful to G. Vignale and I. Tokatly for helping us understand this connection.}.
We emphasize that our results are expected to be exact in the long-wavelength limit of a LFL provided that higher angular momentum Landau parameters ($l\geq2$) are negligible.

\section{Density and current responses}
Any two body operator has a linear expansion in bosonic eigenmodes: 

\begin{eqnarray}
\hat{\mathcal{O}}_{{\bf q}} &=& \int d\theta O({\bf q},\theta) \hat{u}_{{\bf q},\theta} = \sum_\lambda  O_{\lambda,{\bf q}} \hat{\psi}_{\lambda,{\bf q}}.
\end{eqnarray}

\noindent This expansion allows us to quantify the amplitude of the oscillation of any physical quantity in any given eigen-mode and to compute linear response functions. In particular, the density and current operators of the liquid take the following form: $\hat{\rho}_{{\bf q}} = p_{\text{F}}\int d \theta \hat{u}_{{\bf q},\theta}/(2\pi)^2$, ${\bf j}_{{\bf q}} = p_{\text{F}}^2/{m}\int d\theta \hat{u}_{{\bf q},\theta} \hat{{\bf p}}_\theta/(2\pi)^2$, where $m$ is the transport mass that controls the Drude weight. The amplitude of these quantities is found to be:

\begin{eqnarray}
\rho_{\lambda,{\bf q}} &=& \text{sgn}(E_{\lambda}) \frac{p_{\rm F}}{2\pi} \psi^+_{\lambda,0}, \\
{\bf j}_{\lambda,{\bf q}} &=& \text{sgn}(E_{\lambda}) \frac{p_{\rm F}^2}{2\pi m} (\psi^+_{\lambda,1} \hat{{\bf q}}_{||}+\psi^-_{\lambda,1} \hat{{\bf q}}_\perp).
\end{eqnarray}

\noindent Notice that the density and the longitudinal component of the current only have weight in the even parity sector, whereas the transverse component of the current only has weight in the odd parity sector. Therefore the shear sound, which has odd parity, will have purely transverse current oscillations with no accompanying density fluctuations. 

Notably, the imaginary part of the density-density correlation will feature a sharp peak at the energy of the zero-sound mode when it separates from the particle-hole continuum. In a system with $F_0 > 0$ and $F_{l>0} = 0$, the spectral weight of this peak is found to be

\begin{eqnarray}\label{specweight}
w_{\rho \rho, 0} &=& \frac{ p_{\rm F}v_{\rm F}q^2}{8 E_0} \left(1 + \frac{1}{F_0}\right)\left(1 + \frac{1}{2F_0}\right)^{-2}.
\end{eqnarray}

\noindent
This result is consistent with the \textit{f}-sum rule $\frac{n q^2}{m} = \frac{2}{\pi}\int_0^{\infty} d\omega \hspace{1ex}\omega \sum_\lambda w_{\rho \rho, \lambda} \delta (\omega - E_\lambda)$, from which the $F_0 \rightarrow \infty$ limit of the above spectral weight can be approximated. In this limit, the zero sound mode has an energy that is much larger than all other modes so that it exhausts the sum rule, $\frac{n q^2}{m} \simeq \frac{2}{\pi} w_{\rho \rho, 0} E_0 $. Using the electron density $n = \frac{p_\text{F}^2}{4\pi}$ for a circular Fermi surface and writing the effective mass as $m = p_\text{F}/v_\text{F}$, we arrive at $w_{\rho \rho, 0} \simeq \frac{ p_{\rm F}v_{\rm F}q^2}{8 E_0}$, which can also be obtained by directly taking the $F_0 \rightarrow \infty$ limit of Eq.~\eqref{specweight}.

Similarly, the imaginary part of the transverse current-current correlation will feature a sharp peak at the energy of the shear sound mode when it separates from the particle-hole continuum for $F_1>1$ (for details of correlation functions see~\footnotemark[3]). The spectral weight of this peak vanishes as $F_1\rightarrow 1$ and is found to be~\footnotemark[3]:

\begin{eqnarray}
w_{j_{\perp} j_{\perp}} =  \frac{p_{\rm F}^3 v_{\text{F}} q^2}{32 m^2 E_1}\left( 1 - \frac{1}{F_1^2} \right)
\end{eqnarray}

\section{Relation to measurable quantities}
LFLT is parametrized by an infinite number of dimensionless parameters, $\{F_l\}$, whose determination for specific microscopic models can only typically be done approximately. Fortunately, the leading Landau parameters have simple relations to common experimental probes. In particular, $F_1$, controls the ratio of the quasiparticle mass to the transport mass~\cite{Randeria} 

\begin{eqnarray}
\frac{m^*}{m}=1+F_1. 
\end{eqnarray}

Notice that the transport mass only equals the bare mass $m_0$ in Galilean invariant systems~\cite{Baeriswyl,Varma,Okabe}. $m^*$ can be obtained from specific heat measurements, or quantum oscillations, while $m$ can be inferred from the Drude weight, or the London penetration length~\cite{Abrikosov}. Therefore, we expect that in systems where interactions have rendered $m^*\gtrsim 2 m$ ($F_1 > 1$) the shear sound will emerge out of the particle-hole continuum as a sharp excitation, provided that the higher angular momentum Landau parameters ($l\geq 2$) remain small.

\begin{figure}[h]
\includegraphics[scale=0.22]{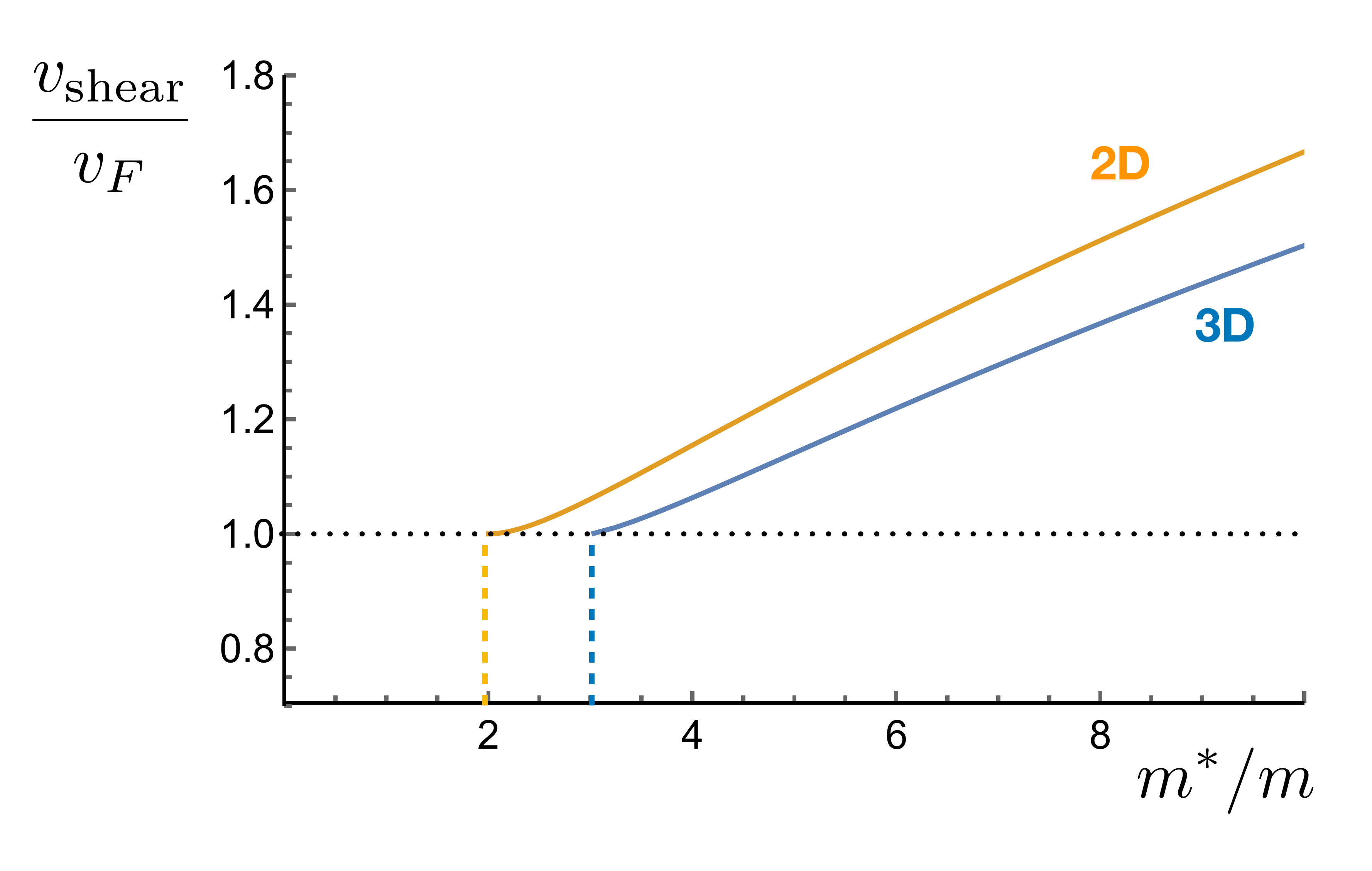}
\caption{\label{2Dvs3D} Comparison of the velocity of shear sound in 2D and 3D as a function of the mass renormalization.}
\end{figure}

\section{Comparison between 2D and 3D}
Although we have focused on two-dimensions similar phenomena can occur in three-dimensions. In fact, the possibility of a shear sound mode in $^3$He was long ago recognized~\cite{Khalat,Fomin,Pines,Baym,Lea}. In 3D a critical Landau parameter $F_1>6$ is required, provided that the higher angular momentum Landau parameters ($l\geq 2$) remain small. To allow a more direct comparison with the two dimensional case, we can relate this to the quasiparticle renormalization, which in 3D is given by~\cite{Pines,Baym}:

\begin{eqnarray}
\frac{m^*}{m}=1+\frac{F_1}{3},   \ {\rm in \ 3D}. 
\end{eqnarray}

\noindent Therefore we can say that in 3D, the shear sound is expected to appear when the quasiparticle mass is renormalised to be three times the transport mass. Figure~\ref{2Dvs3D} shows the behavior of 2D and 3D shear sound modes. Although the Landau parameters of $^3$He are believed to be above the critical value~\cite{Baym}, and there is experimental evidence for it~\cite{Roach}, its quantitative understanding has remained elusive~\cite{Flowers}, largely because it is relatively close to the particle-hole continuum of $^3$He even at largest attainable values of $F_1$~\cite{Baym}. As we will elaborate in the discussion section, strongly interacting two-dimensional metals and helium adsorbed on graphite are promising alternative platforms for the observation of the shear sound mode.

\section{\label{sec:marker}Impact of higher Landau parameters}
So far we have focused on a simple model of shear sound in which higher angular momentum Landau parameters vanish $(F_{l>1}=0)$. In this section we describe the impact of finite higher angular momentum channels on the zero and shear sounds. As it turns out, the qualitative nature of these collective modes remains unchanged even though their solutions are dependent on these higher Landau parameters.

\begin{figure}[b]
\centering
\includegraphics[scale=0.33]{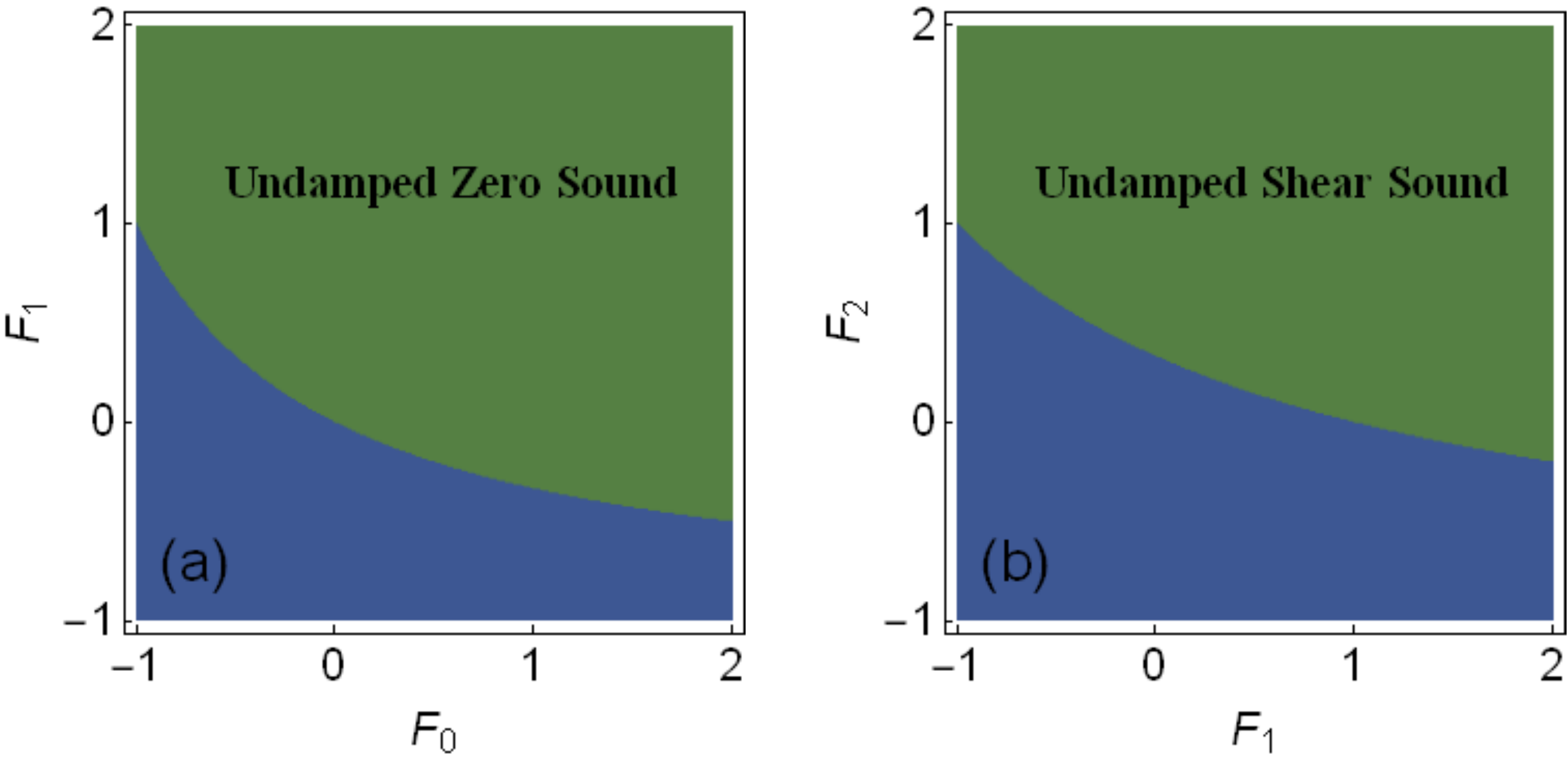}
\caption{\label{soundphasediag}Phase diagrams showing the region (a) in the $F_0-F_1$ parameter space in which the zero sound mode is outside the particle-hole continuum and (b) in the $F_1-F_2$ parameter space in which the shear sound mode is outside the particle-hole continuum.}
\end{figure}

Let us first revisit the even parity sector of the LFL with non-trivial $F_0$ and $F_1$ parameters. As we elaborate in the Supplemental Material~\footnotemark[3], the zero sound mode velocity and wavefunction acquires a dependence on the $F_1$ parameter in addition to the $F_0$ parameter. Figure~\ref{soundphasediag}(a) depicts the modified criterion for the appearance of zero sound for non-zero $F_0$ and $F_1$. Notably, when $F_1 \geq 1$, the zero sound mode is outside of particle-hole continuum for any value of $F_0 \geq -1$ for which the LFL is stable.

From the even to odd sector mapping, an analogous modified criterion for the appearance of shear sound for non-zero $F_1$ and $F_2$ is expected. This is shown in Fig.~\ref{soundphasediag}(b). Likewise, the shear sound remains qualitatively unchanged from the $F_2=0$ case, although its velocity and wave function are now dependent on both the $F_1$ and $F_2$ parameters~\footnotemark[3]. A non-zero $F_2$ alters the minimum value of $F_1$ above which the shear sound mode emerges from the particle-hole continuum. Similarly, when $F_2 \geq 1$, the shear sound mode emerges from the particle-hole continuum for any value of $F_1 \geq -1$ for which the LFL is stable.

A non-zero $F_2$ parameter also modifies the zero sound region in the $F_0-F_1$ space. When its value is sufficiently large, an additional even parity collective mode emerges from the particle-hole continuum which is distinct from the zero sound. More details are discussed in the Supplemental Material~\footnotemark[3]. We expect, however, that these higher angular momentum modes will be harder to realize and probe than the shear sound in typical systems.

\section{Discussion}
We begin by discussing the applicability of our results. For brevity we have focused on spinless fermions but our results apply as well to the case of the symmetric modes of spin unpolarized systems in which spin up and down Fermi surfaces oscillate identically. Also, we have focused on Fermi liquids interacting via short range forces. LFLT in metals requires accounting for the long ranged Coulomb interaction. However, it is not hard to show that the Coulomb interaction modifies only the behavior of the modes in the even sector which involve longitudinal current-density fluctuations, e.g., transforming the zero sound into a plasma mode~\cite{Pines}. Modes in the odd sector, like the shear sound, remain unaltered by the Coulomb interaction because they do not involve charge fluctuations and hence our discussion of these modes is applicable to metals.

\noindent As we have described, we expect that in systems where interactions have rendered $m^*\gtrsim 2 m$ ($F_1 > 1$) the shear sound will emerge out of the particle-hole continuum as a sharp excitation. We suspect that such relatively moderate renormalization should be accessible in a variety of two-dimensional LFLs. For example, in $^3$He films on graphite~\cite{Greywall,Lusher} where $m^*$ diverges on approaching a Mott transition~\cite{Saunders1,Saunders2}. Also in quasi-2D metals near criticality such as the iron based superconductors which have a diverging $m^*$~\cite{Matsuda1,Matsuda2}. It is under debate if both or only one of the masses is enhanced at such critical point~\cite{Chowdhury1,Levchenko,Nomoto,Chowdhury2,Dzero}. The finite and smooth behavior of the residual conductivity near the critical point~\cite{Kasahara} suggests that $m^*$ has greater enhancement than $m$ as required for the appearance of the shear sound~\footnote{Clearly elucidating the ultimate fate of the shear sound in Pnictides would depend on crucial ingredients beyond our model such as the coexistence of electron and hole pockets.}. Additional candidates include quasi-two-dimensional heavy-Fermion materials with large enhancements of the quasiparticle mass~\cite{Haga,Settai,Steglich}, ultracold fermionic gases with enhanced $p$-wave interactions~\cite{Jin,Julienne}, and two-dimensional transition metal dichalcogenides~\cite{TMDC1,TMDC2,TMDC3}.

Finally, we would like to comment on potential experimental probes. One way to study this collective mode is to measure ultra-sound attenuation as attempted in three-dimensional $^3$He~\cite{Roach,Flowers}. Alternatively, in metals, devices like the Corbino viscometer~\cite{Polini} or multi-terminal devices that could generate vorticity of current flow~\cite{Levitov1,Levitov2}, such as those studied in the hydrodynamic approach to electron transport~\cite{Kumar,Philip}, could be used to excite shear sound provided they can be operated in a sufficiently fast dynamical regime to minimize the excitation of particle-hole pairs. It would also be interesting to study the behavior of the shear sound under magnetic fields, which recent studies have incorporated within the bosonization formalism~\cite{Son2,Fradkin3}.

{\bf Acknowledgements}. J. Y. K. is supported by the National Science Scholarship from the Agency for Science, Technology and Research (A*STAR). We are thankful to Dmitrii Maslov, T. Senthil, Dam T. Son, Ady Stern, Maxim Dzero, Atac Imamoglu, Po-Yao Chang, Giovanni Vignale, Ilya Tokatly, Falko Pientka, and Richard Schmidt for valuable discussions.

%

\appendix
\section*{SUPPLEMENTARY MATERIAL}

\section{Derivation of Kinetic Equation Solutions}

We derive in detail the solution to the kinetic equation Eq.~(12) in the main text, which is a recursion relation. We first consider the even parity eigenmodes,

\begin{equation}\label{eq.psieven}
\psi^+_{\lambda ,\theta} = \psi^+ _{\lambda,0}+\sum_{l=1}^\infty 2 \psi^+ _{\lambda,l} \cos (l\theta), 
\end{equation}

of the Landau Fermi Liquid (LFL) with $F_0 \neq 0 $ and $F_{l > 0}= 0$. We will suppress the energy index label $\lambda$ and the parity label $+$ to avoid clutter and reintroduce them at the appropriate discussions that follow. In this case, Eq.~(12) in the main text takes on the simple form

\begin{equation}\label{eq.recrelpsif0}
\psi_{l+2} = 2s\psi_{l+1} - \psi_{l}, \quad l \geq 1
\end{equation}
with the initial conditions,
\begin{subequations}\label{eq.recrelpsif0ic}
\begin{eqnarray}
\psi_{1} &=&  \frac{a_0}{a_1}\frac{s}{\alpha _{1}} \psi_{0} = s \psi_{0} \\
\psi_{2} &=&  \frac{1}{\alpha _{2}}\left( s \psi_{1} - \alpha _0 \psi_{0} \right)=\left(2s^2 -2\alpha _0 \right) \psi_{0},
\end{eqnarray}
\end{subequations}
where $\alpha _l = (1+F_l)/2$ and $\psi_{0}$ is fixed by normalization (see Eq.~\eqref{eq.normcond}). Here we have introduced the reduced energy $s = \frac{E_{\lambda}}{v_{\text{F}} q}$.

Rewriting the above recursion relation Eq.~\eqref{eq.recrelpsif0} as $0 = \sum_{l=3}^{\infty} \left( \psi _{l} - 2s \psi _{l-1} + \psi _{l-2} \right) r^l$ gives rise to the explicit solution for the generating function

\begin{eqnarray}\label{eq.genfunc}
\psi (r) = \sum_{l=0}^{\infty} \psi_{l} r^l = \frac{\psi_{0} f(r)}{(r_+ - r)(r_- - r)},
\end{eqnarray}
where $f(r) = 1 - sr - F_0 r^2$ and $r_\pm = s \pm \sqrt{s^2 -1}$. The wavefunction is then given by the coefficients obtained from expanding the generating function in powers of $r$,

\begin{equation}\label{eq.psin0}
\psi _{l\geq 1} = \frac{\psi_{0}}{r_+-r_-}\left( \frac{f(r_-)}{r_-^{l+1}} - \frac{f(r_+)}{r_+^{l+1}} \right) .
\end{equation}

The behavior of $\psi _{l}$ is different for modes in the particle-hole continuum, $0 < s < 1$, and excitations above the continuum, $s > 1$. When $0< s < 1$, we write $r_{\pm} = e^{\pm i \theta _{E}}$ and find that the wavefunctions (up to a global constant) of such excitations are oscillatory and therefore do not diverge,

\begin{eqnarray}\label{eq.psisolnosc}
\psi _{l\geq 1}^+ 
&=&\frac{\sin(l+1)\theta _{E} -\frac{E}{v_{\rm F}q} \sin l\theta _{E} -F_0\sin (l-1)\theta _{E}}{\sin \theta _{E}}, \nonumber\\
\end{eqnarray}
i.e. Eq.~(15) of the main text. The system therefore supports excitations of any energy $E_{\lambda} < v_{\rm F} q$ and moreover, such excitations are always localized on the Fermi surface and correspond to the quasiparticle excitations of the system.

On the other hand, the wavefunction generally diverges for any arbitrary value $s > 1$ such that solutions do not generally exist. This divergence can be seen from $\lim _{l\rightarrow \infty} \psi _{l} \propto \lim _{n\rightarrow \infty}\frac{f(r_-)}{r_-^{l+1}} \rightarrow \infty$ because $0 < r_- < 1 $ when $s > 1$. However, solutions can exist under specific cases when the numerator of this divergent term vanishes, i.e. when the condition $f(r_-(s)) = 0$ is satisfied. The solution $s_0 > 0$ to this condition is given by Eq.~(16) in the main text,

\begin{equation}\label{eq.v0}
s_0 = \frac{E_0}{v_{\text{F}} q} =\frac{1+F_0}{\sqrt{1+2F_0}}, \quad F_0 > 0,
\end{equation}

where we recover the exact zero sound velocity $v_0 = \frac{E_0}{q}$ obtained from the classical Khalatnikov/Abrikosov approach~\cite{FL0}. The corresponding zero sound wavefunction simplifies to Eq.~(17) of the main text,

\begin{eqnarray}
\psi^+ _{l\geq 1} = \psi^+ _{0} \frac{1+F_0}{(1+2F_0)^{l/2}}, \quad F_0 > 0, \label{eq.psi0wf}
\end{eqnarray}

Unlike modes in the particle-hole continuum, such excitations are always delocalized over the Fermi surface and correspond to the collective modes of the system (see Fig.2 of main text).
The criteria of non-divergence of $\psi _{l}$ explains why at least $F_0 > 0$ is required for the zero sound mode to exist and exactly determines the value of its velocity as a function of $F_0$. 

The normalization constant $\psi^+ _{0}$ is determined from the condition $ \psi_{\lambda,{\bf q},\theta}^\dagger T_{\theta, \theta '}^{-1} \psi_{\lambda',{\bf q},\theta'} = {\rm sgn}(E_\lambda)\delta_{\lambda,\lambda'}$ for the case when $\lambda = \lambda ' = s_0$,


\begin{eqnarray}
1  &=& \psi^+ _{s_0, \theta} \mathcal{T}^{-1}_{\theta, \theta '}\psi^+ _{s_0, \theta'} \nonumber \\
&=& \frac{p_{\rm F}}{(2\pi)^2 q}\sum _{l,m = 0} ^{\infty} a_l a_m \psi^+ _{s_0, l} \psi^+ _{s_0, m} \gamma _{lm}, \label{eq.normcond} \\
&& \gamma _{lm} = \int d\theta \frac{\cos (l \theta) \cos (m \theta)}{\cos (\theta)},
\end{eqnarray}
where $a_0 = 1$ and $a_{l>0} = 2$. It can be shown that the only non-trivial values of $\gamma _{lm}$ are
\begin{equation}
\gamma _{2n+1,2m} = \gamma _{2m,2n+1} = 2\pi (-1)^{n+m}, \quad 0 \leq m \leq n \in \mathbb{Z},
\end{equation}
Substituting Eq.~\eqref{eq.psi0wf} into Eq.~\eqref{eq.normcond} and evaluating the sum explicitly, one eventually arrives at the normalization constant given by Eq.~(18) in the main text,

\begin{equation}
\psi^+ _{0} = \left(\frac{2\pi q}{p_{\rm F}} \frac{v_{\text{F}} q}{E_0} \frac{F_0 (1 + F_0)}{(1+2F_0)^2}\right)^{1/2}, \quad F_0 > 0.
\end{equation}



For a LFL with non-trivial $F_{l < 2}$ and $F_{l \geq 2} = 0$, the steps above can be repeated for the odd parity eigenmodes,

\begin{equation}
\psi^{-}_{\lambda,\theta} = \sum_{l=1}^\infty 2 \psi^-_{\lambda,l} \sin (l\theta).
\end{equation}

In this case, Eq.~(12) in the main text takes an analogous form 

\begin{equation}\label{eq.recrelpsif1}
\psi_{n+3} = 2s\psi_{n+2} - \psi_{n+1}, \quad n \geq 1
\end{equation}
with the initial conditions,
\begin{subequations}\label{eq.recrelpsif1ic}
\begin{eqnarray}
\psi_{2} &=&  \frac{a_1}{a_2}\frac{s}{\alpha _{2}} \psi_{1} = 2s \psi_{1} \\
\psi_{3} &=&  \frac{1}{\alpha _{3}}\left( s \psi_{2} - \alpha _1 \psi_{1} \right)=\left(4s^2 -2\alpha _1 \right) \psi_{1}.
\end{eqnarray}
\end{subequations}

The above recursion relation can be solved in a similar fashion to the even parity case to obtain the odd parity collective mode solution, i.e. the \textit{shear sound} mode. The velocity $v_1 = s_1 v_{\text{F}}$ and wavefunction $\psi^{-}$ of the shear sound are respectively,

\begin{eqnarray}
s_1 &=& \frac{E_1}{v_{\text{F}} q} =\frac{1+F_1}{2\sqrt{F_1}}, \quad F_1 > 1, \label{eq.v1} \\
\psi^- _{l\geq 2} &=&  \psi^{-}_{1} \frac{F_1+1}{F_1^{l/2}}, \quad F_1 > 1, \label{eq.psi1wf} 
\end{eqnarray}
as per Eq.~(19) and Eq.~(20) in the main text. 
The normalization condition for the shear sound reads,

\begin{eqnarray}
1  &=& \psi^- _{s_1,\theta} \mathcal{T}^{-1}_{\theta, \theta '}\psi^- _{s_1,\theta}\nonumber \\
&=& \frac{p_{\rm F}}{(2\pi)^2 q}\sum _{l,m = 1} ^{\infty} a_l a_m \psi^- _{s_1, l} \psi^- _{s_1, m} \tilde{\gamma} _{lm}, \label{eq.normcondodd} \\
\tilde{\gamma} _{lm} &=& \int d\theta \frac{\sin (l \theta) \sin (m \theta)}{\cos (\theta)}
\end{eqnarray}
It can be shown that the only non-trivial values of $\tilde{\gamma} _{lm}$ are
\begin{equation}
\tilde{\gamma} _{2n,2m-1} = \tilde{\gamma} _{2m-1,2n} = 2\pi (-1)^{n+m}, \quad 1 \leq m \leq n \in \mathbb{Z}
\end{equation}

Substituting Eq.~\eqref{eq.psi1wf} into Eq.~\eqref{eq.normcondodd} and evaluating the sum explicitly, one eventually arrives at the normalization constant given by Eq.~(20) in the main text,

\begin{equation}
\psi^- _{1} = \left(\frac{\pi q}{8 p_{\rm F}} \frac{v_{\text{F}} q}{E_1} \left( 1 - \frac{1}{F_1^2} \right)\right)^{1/2}.
\end{equation}

\section{Solutions involving higher Landau parameters}

In the previous section, the analytic collective mode solutions were derived for the cases which had dependence on only one Landau parameter. While the introduction of higher Landau parameters necessarily complicates the solution, the general framework introduced in the previous section can still be used to obtain the constrain equation satisfied by the collective mode velocities, i.e. the criteria of non-divergence of the wavefunction components.

In the tight-binding model picture (Section III of main text), the effect of introducing higher Landau parameters is to extend the boundary deeper into chain. Equivalently, more initial conditions are generated. These recursion relations, together with the initial conditions, can always be used to construct generating functions of the form given by Eq.~\eqref{eq.genfunc}. Specifically, the polynomial $f(r)$ gets modified by the higher Landau parameters by generating coefficients of the higher powers of $r$. For instance, for non-zero parameters $\left\lbrace F_0, F_1, F_2 \right\rbrace$, the polynomial $f(r)$ for the even parity solution reads,

\begin{eqnarray}
f(r) &=& 1 + \sum _{l=1} f_l r^l, \\
f_1 &=& -2s + \frac{s}{2\alpha _1}, \quad
f_2 = 1 - \frac{\alpha _0}{\alpha _2} + \frac{s^2}{\alpha _1}\left( \frac{1}{2\alpha _2} - 1 \right), \nonumber \\
f_3 &=& -s + \frac{s}{2\alpha _1}, \quad
f_4 = 2\alpha _0 - \frac{\alpha _0}{\alpha _2} + \frac{s^2}{\alpha _1}\left(\frac{1}{2\alpha _2} -1 \right),\nonumber \\
f_{l \geq 5} &=& 0. \nonumber
\end{eqnarray}
Regardless of the actual form of $f(r)$, the constrain equation is always the same,
\begin{equation}
f(r_-) = 0, \quad r_- = s - \sqrt{s^2-1}.
\end{equation}
The solutions $\lbrace s_i > 1\rbrace$ to the above constrain equation corresponds to the collective mode velocities, $v_i = s_i v_{\text{F}}$. The corresponding wavefunctions to these collective modes can be obtained by substituting their respective velocities into the general wavefunction expression.

\begin{figure}[t]
\centering
\includegraphics[scale=0.6]{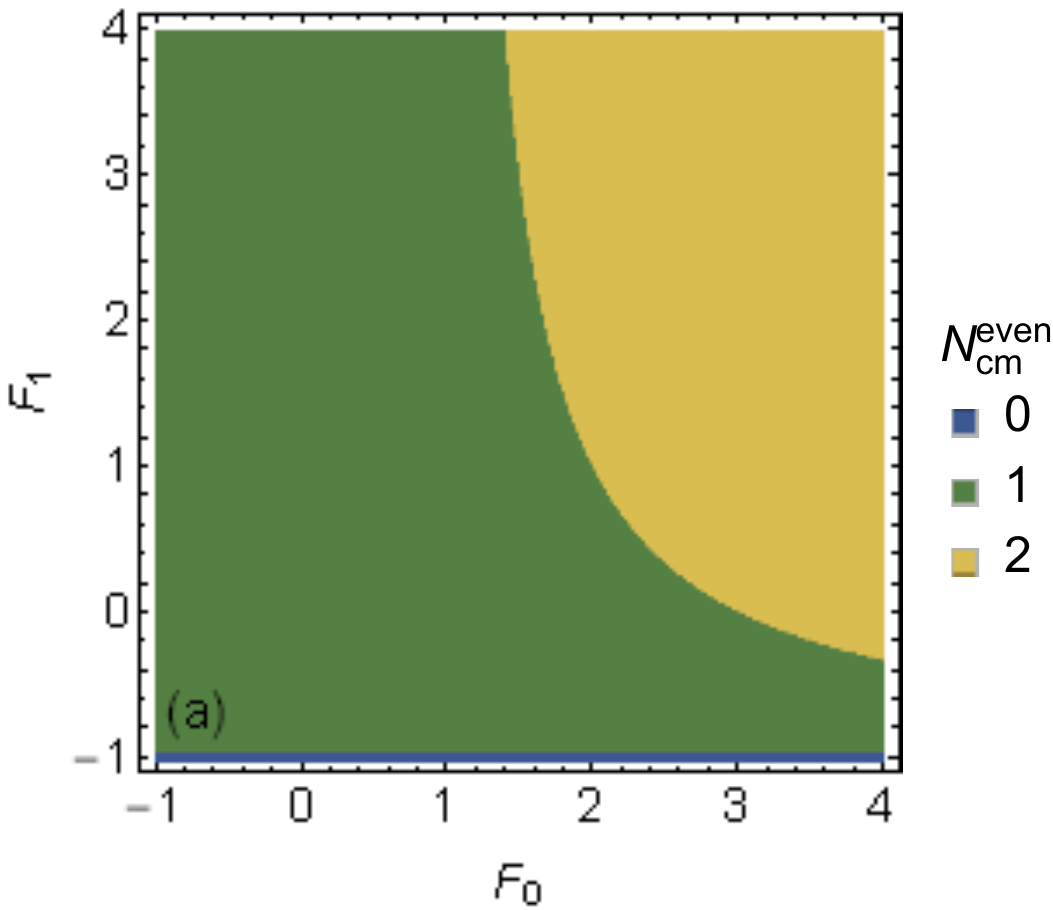} \\ \vspace{1em}
\includegraphics[scale=0.295]{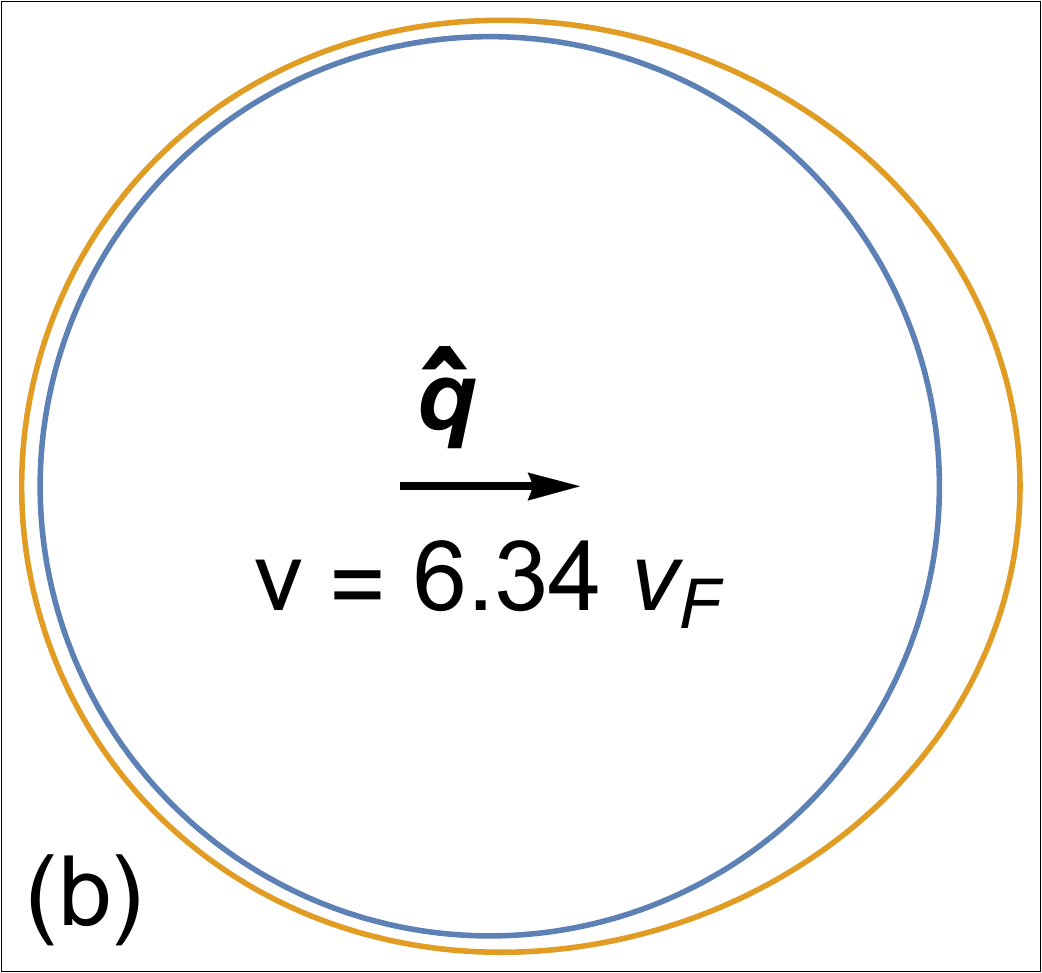} \hspace{3em}
\includegraphics[scale=0.275]{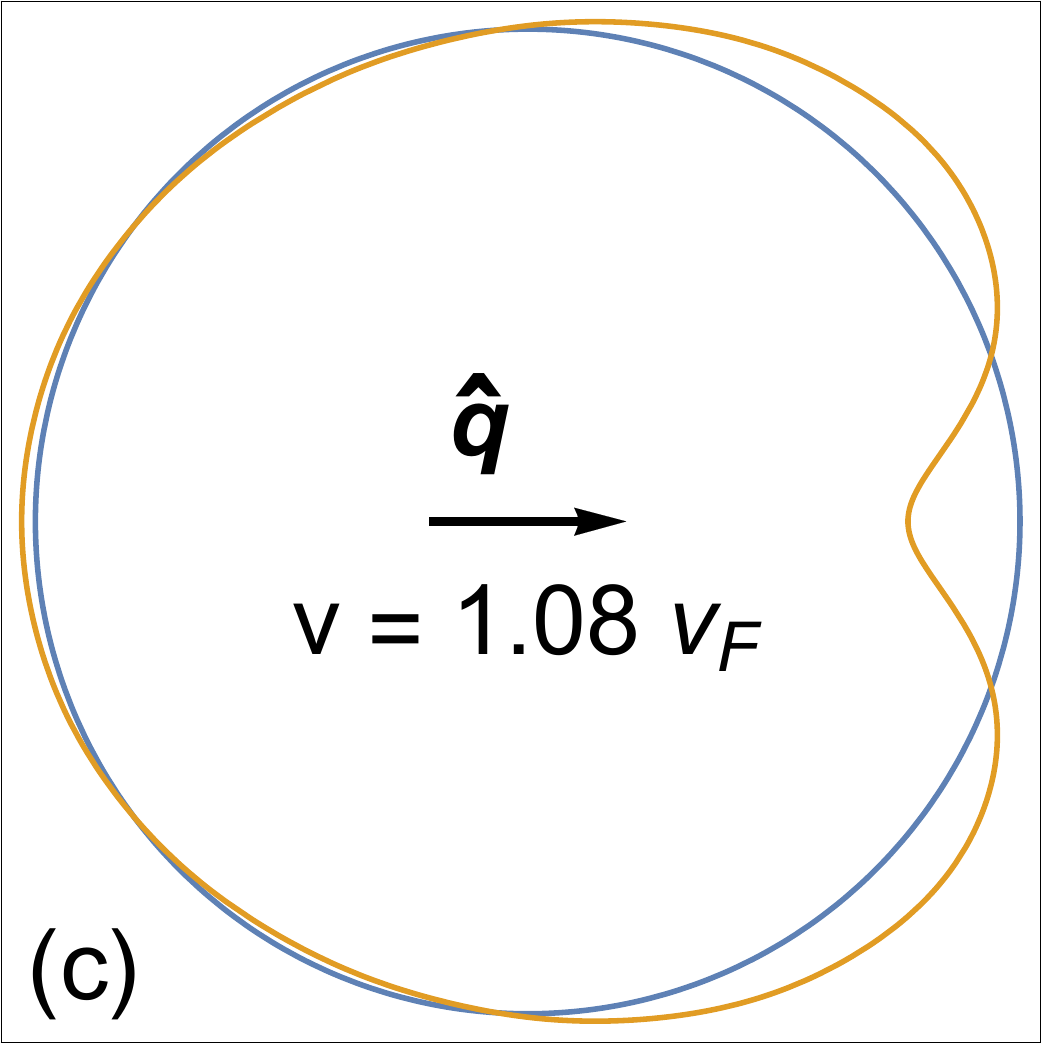}
\caption{\label{F0F1ephaseF2p}(a) Phase diagram showing $N_{\rm cm}^{\rm even}$, the number of even parity collective modes outside the particle-hole continuum, i.e. with energy $E > v_{\rm F}q$ for $F_2 = 3$. Fermi surface deformations of collective modes outside the particle-hole continuum for $F_0 = 7, F_1 = 7$ and $F_2 = 3$ with velocity (b) $v = 6.34 v_{\rm F}$ and (c) $v =  1.08 v_{\rm F}$.}
\end{figure}

As was discussed in the main text, the LFL with non-zero parameters $\left\lbrace F_0, F_1 \right\rbrace$ allow for new conditions in which the zero sound mode emerges out of the particle-hole continuum. Introducing a sufficiently large $F_2$ parameter additionally allows for more than one even parity collective mode solution, i.e. an additional higher angular momentum collective mode in addition to the zero sound. In Fig.~\ref{F0F1ephaseF2p}, we show the region in the $F_0-F_1$ space for which $N_{\rm cm}^{\rm even} = 0,1,2$ number of even parity collective mode solutions exist when $F_2 = 3$. When two collective mode solutions exist, the one with the larger velocity corresponds to the zero sound while the one with the lower velocity corresponds to a higher angular momentum mode, which can be seen by comparing their corresponding Fermi surface deformations shown in Fig.~\ref{F0F1ephaseF2p}(b) and (c) with that of the zero sound shown in Fig.~2(a) of the main text.

This preliminary investigation suggests that there may be many additional collective modes that can emerge from the particle-hole continuum when higher angular momentum Landau parameters are introduced. From the above figures, we expect the wavefunctions of these modes will be qualitatively distinct from that of the zero sound. From the even to odd sector mapping, introducing higher angular momentum Landau parameters will necessarily lead to the emergence of additional odd parity collective modes distinct from the shear sound as well. Further investigations beyond the scope of this paper will be required to determine the properties of these additional collective modes and their relation to the zero and shear sound mode.

\section{Derivation of Density and Current\\Response Functions}

In this section, we provide a detailed derivation of the response functions given in the main text. The operator corresponding to an observable $\mathcal{O}({\bf q}) = \left\langle \hat{\mathcal{O}}_{{\bf q}} \right\rangle $ can be explicitly expressed in terms of its quantum eigenmodes,

\begin{eqnarray}
\hat{\mathcal{O}}_{{\bf q}} &=& \int d\theta O({\bf q},\theta) \hat{u}({\bf q},\theta) = \sum_\lambda  O_{\lambda,{\bf q}} \hat{\psi}_{\lambda,{\bf q}}, \\
O_{\lambda,{\bf q}} &=& \text{sgn}(E_\lambda)\int d\theta O({\bf q},\theta)\psi _{\lambda,{\bf q},\theta}.
\end{eqnarray}

This operator evolves as $\hat{\mathcal{O}}_{{\bf q}} (t)= \sum_{\lambda} O_{\lambda,{\bf q}} \hat{\psi}_{\lambda,{\bf q}} ({\bf q})e^{-iE_{\lambda} t}$, which simplifies the computation of response functions, $\chi _{AB} ({\bf q},t) = -i \Theta (t) \left\langle \left[ \hat{A}_{{\bf q}}(t) ,\hat{B}_{-{\bf q}} \right] \right\rangle $. It can be shown that its imaginary part in frequency domain reads,

\begin{equation}\label{eq.chiomegaIm}
\text{Im} \chi _{AB}({\bf q},\omega )= -\pi \mathcal{A} \sum_{\lambda} \text{sgn}(E_{\lambda})  A _{\lambda ,{\bf q}} B _{\lambda ,-{\bf q}} \delta(\omega- E_{\lambda}).
\end{equation}

The density $\rho ({\bf x})$ and current ${\bf j}({\bf x})$ of a LFL with a distribution function $ n(t,{\bf x},{\bf p}) = \Theta \left( p_{\text{F}}(t,{\bf x},\theta) - p \right)$ are

\begin{eqnarray}
\rho ({\bf x}) &=& \int \frac{d^2{\bf p}}{(2\pi)^2}n({\bf x},{\bf p}) = \rho_0 + p_{\text{F}}\int \frac{d \theta}{(2\pi)^2}u({\bf x},\theta), \nonumber \\
{\bf j}({\bf x}) &=& \left( j_{\parallel}({\bf x}), j_{\perp}({\bf x}) \right) = \frac{p^2_{\text{F}}}{m}\int\frac{d\theta}{(2\pi)^2}u({\bf x},\theta ) \hat{{\bf p}}_\theta. \nonumber 
\end{eqnarray}
to first order in $u({\bf x},\theta)$. We define the current via its transport mass $m$ and explicitly separate the current into its longitudinal $j_{||}$ and transverse $j_{\perp}$ components. The corresponding operator coefficients are
\begin{eqnarray}
\rho_{\lambda,{\bf q}} &=& \text{sgn}(E_{\lambda}) \frac{p_{\rm F}}{2\pi} \psi^+_{\lambda,0}, \\
j_{\parallel,\lambda,{\bf q}} &=& \text{sgn}(E_{\lambda}) \frac{p^2_{\rm F}}{2\pi m} \psi^+_{\lambda,1}, \\
j_{\perp,\lambda,{\bf q}} &=& \text{sgn}(E_{\lambda}) \frac{p^2_{\rm F}}{2\pi m} \psi^-_{\lambda,1},
\end{eqnarray}

From Eq.~\eqref{eq.chiomegaIm}, we can simply read off the density-density and current-current correlation functions,

\begin{eqnarray}
\text{Im} \chi _{\rho \rho}({\bf q},\omega )&=& - \mathcal{A} \frac{p_{\rm F}^2}{4\pi } \sum_{i} \left|\psi^+_{i,0}\right|^2 \text{sgn}(E_i) \delta(\omega- E^+_i),  \nonumber \\ \\
\text{Im} \chi _{j_{\parallel} j_{\parallel}}({\bf q},\omega ) &=& - \mathcal{A} \frac{ p_{\rm F}^4}{4\pi m^2 } \sum_{i} \left|\psi^+_{i,1}\right|^2 \text{sgn}(E_i) \delta(\omega- E^+_i), \nonumber \\ \\
\text{Im} \chi _{j_{\perp} j_{\perp}}({\bf q},\omega ) &=& - \mathcal{A} \frac{ p_{\rm F}^4}{4\pi m^2 } \sum_{i} \left|\psi^-_{i,1}\right|^2 \text{sgn}(E_i) \delta(\omega- E^-_i).  \nonumber \\
\end{eqnarray}

In particular, for a system with only non-trivial $F_0$, the density-density and longitudinal current-current correlation functions will exhibit a sharp peak at the zero-sound energy $E_0 = q v_\text{F} \frac{1+F_0}{\sqrt{1+2F_0}}$ with the following spectral weights,

\begin{eqnarray}
w_{\rho \rho, 0} &=& \mathcal{A}\frac{ p_{\rm F}q}{8} \frac{1}{s_0} \left(1 + \frac{1}{F_0}\right)\left(1 + \frac{1}{2F_0}\right)^{-2}, \\
w_{j_{\parallel} j_{\parallel}, 0} &=& \mathcal{A}\frac{ p_{\rm F}^3q}{8m^2} s_0 \left(1 + \frac{1}{F_0}\right)\left(1 + \frac{1}{2F_0}\right)^{-2}.
\end{eqnarray}	

On the other hand, for a system with non-trivial $F_0$ and $F_1$, the transverse current-current correlation function will exhibit a sharp peak at the shear sound energy $E_1 = q v_\text{F} \frac{1+F_1}{2\sqrt{F_1}}$ with spectral weight

\begin{eqnarray}
w_{j_{\perp} j_{\perp}, 1} &=&  \mathcal{A}\frac{ p_{\rm F}^3 q}{32m^2} \frac{1}{s_1} \left( 1 - \frac{1}{F_1^2} \right) \\
&=& \left.\frac{1}{4}\frac{p_{\text{F}}^2}{m^2} w_{\rho \rho, 0} \right|_{F_0 \rightarrow \frac{1}{2}(F_1-1)},
\end{eqnarray}

where in the last line, we found an interesting mapping between the even and odd sector spectral weights between these correlation functions.

\end{document}